\documentclass[12pt,preprint]{aastex}

\slugcomment{Not to appear in Nonlearned J., 45.}

\shorttitle{3 poststarburst galaxies} \shortauthors{Yufeng Mao et
al.}

\usepackage{amssymb}
\usepackage{pifont}
\usepackage{amsmath}

\begin{document}


\title{Extending the Eigenvector 1 Space to the Optical Variability of Quasars}


\author{Yufeng Mao\altaffilmark{1,2}, Jing Wang\altaffilmark{1}, \& Jianyan Wei\altaffilmark{1}}
\altaffiltext{1}{National Astronomical Observatories, Chinese
Academy of Sciences, Beijing 100012, China; myf@bao.ac.cn}
\altaffiltext{2}{Graduate University of Chinese Academy of Sciences,
Beijing, P.R.China}


\begin{abstract}

We introduce a new physical parameter, the optical variability
amplitude, to the well-established Eigenvector 1 space of quasars
and test a sample of long-term B-band light curves of 42 PG quasars
monitored by Giveon et al. (1999). We find that the optical
variability amplitude strongly correlates with the intensity ratio
of Fe II to H$\beta$, H$\beta$ width and peak luminosity at
5007$\rm{\AA}$. We briefly discuss the physical meaning of our
findings and suggest that the Eddington ratio may be a key factor in
determining a quasar's variability.

\end{abstract}


\keywords{galaxies: active --- quasars: general}



\section{Introduction}

During the last few decades, quasars have been monitored in
multi-wavelength observations, from radio to X-rays, by many
research programs. Our understanding of the central engine of
quasars has been greatly improved by the correlations found between
the variability properties and other observational parameters (e.g.,
luminosity, redshift, rest-wavelength, timescales and emission-line
width). An anticorrelation between the amplitude of variability and
luminosity was reported by Pica \& Smith (1983), and confirmed by
many subsequent studies (e.g., Cristiani et al. 1990; Hook et al.
1994; Cristiani et al. 1997; Giveon et al. 1999 (hereafter G99);
Garcia et al. 1999; Vanden Berk et al. 2004), although there have
also been reports to the contrary (e.g., Trevese et al. 1989;
Giallongo et al. 1991; Cimatti et al. 1993). The relationship
between the variability amplitude and redshift was discussed in many
studies. Some authors found that the variability amplitude is
anticorrelated with redshift (e.g., Barbieri et al. 1983; Cristiani
et al. 1990; Hook et al. 1994; Cristiani et al. 1996), while others
reported an opposite trend (e.g., Giallongo et al. 1991; Treverse et
al. 1994; Cid Fernandes et al. 1996). Moreover, it was found that
the optical spectra usually become harder as quasars turn brighter
(Cutri et al. 1985; G99). Kollatschny et al. (2006) examined the
variability properties of a sample of 10 Palomar-Green quasars with
the line width of H$\beta$ larger than 5000 km $s^{-1}$. They found
a marginal correlation between the optical continuum variability
amplitude and H$\beta$ line width, which provides useful information
for understanding the structure of BLR.

On the other hand, Boroson \& Green (1992; hereafter BG92) examined
a sample of 87 bright low redshift PG quasars and found that most of
the variance is connected to two sets of correlations, which were
then defined as Eigenvector 1 (hereafter E1) space and Eigenvector 2
space. The E1 space is dominated by the strong anticorrelation
between the optical FeII and [OIII], and Eigenvector 2 by the
correlation between the optical luminosity and HeII equivalent width
(see Sulentic et al. 2000a for a review). By calculating the virial
black hole mass ($\rm{M_{BH}}$) using the well established empirical
$R-L$ relationship (e.g., Kaspi et al. 2000), Boroson (2002)
suggested that the E1 space is dominantly driven by the Eddington
ratio L/$\rm{L_{Edd}}$, and the Eigenvector 2 by $\rm{M_{BH}}$ (see
also in Sulentic et al. 2000b).

In this paper, we will investigate whether the quasar's optical
variability amplitude is related with the E1 space. We test a sample
of 42 PG quasars. The paper is structured as follows. The sample
selection and our analysis are described in Section 2. The results
are presented in Section 3, and discussions in Section 4. Throughout
this paper, the $\Lambda$CDM cosmology (Spergel et al. 2003) with
$\Omega_{m}=0.3$, $\Omega_{\Lambda}=0.7$, and ${\rm
H_{0}=70~km~s^{-1}~Mpc^{-1}}$ are adopted.


\section{Sample and Analysis}
\label{sect:Sample}

We searched the literature for suitable results of quasars' optical
variability. In particular, the light curves should be well sampled
in the temporal domain, with adequate total observation time and
sampling interval. It is found that the G99 subset of the PG quasars
is well suited for our purpose, because the optically selected PG
quasars are not only nearly statistically complete but also studied
comprehensively in multi-wavelength observations. G99 monitored 42
nearby ($z<0.4$), bright ($B<16$mag) PG quasars in the B and R bands
for a seven-year period at Wise Observatory. The typical temporal
sampling interval was 40 days, and the objects were observed at
30-60 epochs with a photometry uncertainty of $\sim0.01$mag. All the
objects showed intrinsic rms variability amplitudes of 5\%
$<\sigma_B<$34\% and 4\% $<\sigma_R<$26\% .

Our sample is listed in Table 1. Figure 1a shows the distribution of
the redshifts for the sample listed in G99, see also Column (2) of
Table 1. The redshifts of these PG quasars are mainly less than 0.2.
Column (5) lists the variability amplitude in magnitude for each
object. Furthermore, the variability amplitude can be statistically
estimated in various ways. We refer the readers to G99 for a brief
comment on these different methods. In the current study, the
variability is defined as the median value of all possible magnitude
difference of a light curve, simply because the median value is a
relatively robust estimation, i.e., not strongly affected by the
outliers (e.g., Hook et al. 1994, Netzer et al. 1996). Only the
variability in the B band is considered in the subsequent analysis
since the variability is more significant in the B band than in the
R band. G99 examined the correlation between the variability
amplitude defined by magnitude and quasar spectral properties
defining the E1. However, no significant correlations were found by
them. It should be pointed out that the variability amplitude
defined in magnitude represents a \it relative \rm change in
luminosity. For a constant change defined in luminosity, a small
(large) change in magnitude could be simply caused if the object is
(less) luminous. In order to overcome this problem, the median of
absolute change in luminosity is used in this paper, that is
\begin{equation}
\Delta L=L_{bol} \times (1-10^{-0.4\times |median(\Delta B)|})
\end{equation}
where $median(\Delta B)$ is the median of variability in the B band
and $\rm{L_{bol}}$ the fiducial luminosity of the quasar. Note that
$\Delta L$ is always positive according to this definition. $\Delta
L$ represents the characteristic variability of each quasar in the
absolute luminosity change, which directly reflects the absolute
change of the amount of the fueling gas. Regarding two quasars with
the same change in the $median(\Delta B)$, the one with a larger
bolometric luminosity would have a larger $\Delta L$.

The bolometric luminosity can be estimated in two ways. First, we
use an widely accepted empirical relationship
\begin{equation}
L_{\rm{bol}}=9\lambda L_\lambda(5100\AA)
\end{equation}
as given by Kaspi et al. (2000), where the luminosities at the
rest-frame wavelength 5100\AA\ are adopted from Vestergaard \&
Peterson (2006). In another way, we also calculate the
$L_{\rm{bol}}$ from the median apparent magnitudes in the B band
given by G99, using the formula
\begin{equation} log(L_{bol}
/\upsilon _{B} L_{\upsilon _{B}}) = 0.80 - 0.067\mathcal {L} + 0.017
\mathcal {L} ^{2} - 0.0023 \mathcal {L} ^{3}
\end{equation}
of Marconi et al. (2004), where $\mathcal {L}$ = (log $\rm{L_{bol}}$
-12). A comparison of between the values of $\rm{L_{bol}}$ obtained
from Equation (2) and those from Equation (3) is made (see Figure
2). There is a strong correlation (with slope $\sim$ 1) between the
two sets of $\rm{L_{bol}}$, indicating that the two independent
measurements are highly consistent with each other. So we adopt the
value of $\rm{L_{bol}}$ obtained from Equation (2), and then
calculate the $\Delta L$ using Equation (1). The calculated $\Delta
L$ in the logarithm is shown in Column (6) of Table 1 for each
quasar.

\section {Results}

The main goal of the present paper is to investigate whether the
variability amplitude of quasars is related to E1 space. The E1
space is dominated by significant correlations between RFe
(=FeII/H$\beta$), FWHM(H$\beta$), and [OIII] strength, and has been
discussed by many authors (e.g. BG92; Xu et al. 2003; Grupe  2004;
Sulentic et al. 2000a). We list the E1 parameters of our PG quasar
sample in Table 1. Column (7) gives the FWHM of the broad component
of H$\beta$, Column (8) the ratio of the peak hight of [OIII]
$\lambda$5007 to that of H$\beta$ (Peak $\lambda$5007), Column (9)
the ratio of the flux of FeII integrated in the rest frame
wavelength range from $\lambda$4434 to $\lambda$4684 to that of
H$\beta$ (RFe), Column (10) the logarithm of R, i.e., the ratio of
radio flux at 6 cm to optical flux density. All of these parameters
are adopted from BG92.

We then investigate the correlations between $\Delta L$ and E1
parameters in our sample. Figure 3 displays the correlations of
$\Delta L$ versus FWHM(H$\beta$), peak $\lambda$5007, RFe, and the
radio loudness $\log R$, respectively. The spearman rank-ordered
correlation coefficients $r_{s}$ of the four correlations are listed
in Table 2, where $P_s$ is the probability of null correlation.
Figure 3a shows a significant correlation between the $\Delta L$ and
FWHM(H$\beta$). A spearman rank-ordered analysis yields a
correlation coefficient $r_s$=0.450 with a significance level
$P_s$=0.004. This means that the quasars with larger widths of
H$\beta$ would have larger changes in luminosity. We also find a
correlation between $\Delta L$ and peak $\lambda$5007 (Fig.3b,
$r_s$=0.445, $P_s$=0.004). The anti-correlation between $\Delta L$
and RFe is plotted in Fig.3c. The calculated correlation coefficient
is $r_s$=-0.441, and the significance level $P_s$=0.005. Kollatschny
et al. (2006) did not find a significant correlation between the
continuum variability amplitude at 5100\AA\ and radio power at 5GHz
in their sample. In current studies, a correlation ($r_s$=0.476,
$P_s$=0.0023) is identified between the radio loudness $\log R$ and
$\Delta L$, and is shown in Fig.3d. Since $\log R=1$ is widely used
to separate the radio-loud and radio-quiet quasars (Kellermann et
al. 1989), the diagram shows that all the radio-loud quasars ($\log
R>1$) have large optical variability amplitudes ($\log(\Delta
L)>45$), although the radio-quiet quasars ($\log R<1$) are nearly
evenly distributed in terms of variability amplitude. The fact that
radio-loud quasars have large variability amplitudes implies that
the optical continuum of radio-loud quasars is contaminated by the
high energy tail of their radio emissions, which boosts the
variability amplitude because of the beam effect of the jets.

The correlations found above suggest that the E1 space could be
extended to include the variability amplitude. This hypothesis can
be verified by a principal component analysis (PCA) of our sample.
The PCA is performed using the following 11 parameters, which are:
$\log R$, the equivalent width of H$\beta$, R($\lambda$5007),
R($\lambda$4686), RFe, Peak($\lambda$5007), FWHM(H$\beta$), H$\beta$
shift, H$\beta$ shape, H$\beta$ asymmetry, and $\log (\Delta L)$,
each potentially providing unique information. Except for $\log
(\Delta L)$, the former ten parameters are directly collected from
BG92. We refer the reader to BG92 for the definitions of these
parameters. The PCA results are presented in Table 3, which lists
the first four most significant eigenvectors. The second row shows
the cumulative percentage of the variance. One can see that the
first four eigenvectors together account for more than 70 percent of
the variance, and that the first principal component dominates the
observed properties of quasars. Similar to BG92, the E1 is dominated
by the anticorrelation between the strength of FeII and Peak
($\lambda$5007). In addition, our E1 is strongly effected by $\log
(\Delta L)$. It is clear that $\log (\Delta L)$ has a projection of
0.54 on E1, and 0.57 on the Eigenvector 2. Although at first sight,
the projection on Eigenvector 2 is larger than on E1, taking the
larger cumulative percentage of the E1 into account, we conclude
that the E1 space can be extended to $\log (\Delta L)$.

\section {Discussion}
\label{sect:discussion}

\subsection{Variability vs. $\rm{M_{BH}}$ and Eddington ratio}
After extending the E1 space to the variability in amplitude, the
dominant physical parameters are discussed in this section. Both of
the black hole mass ($\rm{M_{BH}}$) and Eddington ratio
(L/$\rm{L_{Edd}}$) are believed to be the main parameters governing
the observed properties in quasars. However, the luminosity
variability defined by Equation (1) cannot be used to correlate with
$\rm{M_{BH}}$ and L/$\rm{L_{Edd}}$ because both $\rm{M_{BH}}$ and
L/$\rm{L_{Edd}}$ are estimated in terms of the continuum luminosity
which is used to define the variability in luminosity. The
variability in magnitude $median(\Delta B)$ is therefore used
instead in the subsequent analysis. All $\rm{M_{BH}}$ values are
adopted from Vestergaard \& Peterson (2006). The distribution of
$\rm{M_{BH}}$ is shown in Figure 1b. One can see that the majority
of log($\rm{M_{BH}}$/$\rm{M_{\odot}}$) lie between 7.5 and 9.5.
$\rm{L_{bol}}$ is then estimated using the Equation (2). The values
of $\rm{M_{BH}}$ and L/$\rm{L_{Edd}}$ are listed in Column (3) and
(4) in Table 1, respectively.

The relation between the variability amplitude and $\rm{M_{BH}}$ or
L/$\rm{L_{Edd}}$ has been discussed recently. Contradictory results
were, however, obtained by different authors. Wold et al. (2007)
examined the relation between the quasar variability and black hole
mass by studying the optical variability of $\sim$100 quasars
monitered by QUEST1 survey (Rengstorf et al. 2004). However, a
correlation between the R-band variability and $\rm{M_{BH}}$ was
marginally obtained only when $\rm{M_{BH}}$ was averaged in several
bins. Furthermore, they did not detect such a relation in their PG
quasar sample. In contrast, Wilhite et al. (2008) found a
correlation between the variability and $\rm{M_{BH}}$ in a sample of
$\sim$2500 quasars selected from the SDSS. In addition, they
reproduced the well-known anti-correlation between the variability
and luminosity. By combining the two relations, they suspected that
L/$\rm{L_{Edd}}$ is a possible driver for quasar variability in an
indirect way.

As $\rm{M_{BH}}$ $\propto$ $\rm{(FWHM)^2}$, a possible way to test
the correlation between variability and $\rm{M_{BH}}$ is to search
for the correlation between the variability and line width, both of
which are independent observational parameters. The relation between
the variability amplitude and line width was discussed in several
papers. G99 found a marginal correlation between the variability
amplitude (defined in magnitude) and width of the H$\beta$ emission
line. Their possible, but unlikely explanation, for this trend is
the contributions of the emission lines to the broad-band emission.
Kollatschny et al. (2006) confirmed the results of G99 in a sample
of 43 galaxies. In the above analysis, we find a significant
correlation between $\log(\Delta L)$ and the FWHM of H$\beta$ (Fig.
3a).

The median($\Delta$B) is plotted against $\rm{M_{BH}}$ in Figure 4a.
No significant correlation is, however, found between these two
parameters ($r_s = 0.124$, $P_s = 0.4267$). Wold et al. (2007) did
not detect a correlation between variability and $\rm{M_{BH}}$ in
their PG quasar sub-sample, and our result confirms their
conclusion. However, the current results make it difficult to
understand why the correlation between the magnitude variability and
the $\rm{M_{BH}}$ is not as good as expected.

It is now generally believed that the E1 spaces is likely driven by
L/$\rm{L_{Edd}}$ (e.g., Boroson 2002). The relation between the
magnitude variability and L/$\rm{L_{Edd}}$ is directly examined in
Figure 4b. We find a significant anti-correlation between the
median($\Delta$B) and L/$\rm{L_{Edd}}$ ($r_s = -0.368$, $P_s =
0.0012$). Although our result agrees with Wilhite et al. (2008),
caution must be made when explaining the
median($\Delta$B)-L/$\rm{L_{Edd}}$ correlation. Taking two quasars
with the same $\rm{M_{BH}}$, the more luminous one would have
smaller variability in magnitude for a given change in luminosity
(accretion rate). Meanwhile, the more luminous quasar would have a
larger L/$\rm{L_{Edd}}$. That means that the
median($\Delta$B)-L/$\rm{L_{Edd}}$ correlation might be caused by an
intrinsic relation in mathematics (i.e., the definition of
magnitude) rather than in physics. A sample of light curves defined
in flux or luminosity is therefore required to test the underlying
physics of the relation.

\subsection{Implications on variability mechanisms}

Several theoretical models have been proposed as mechanisms for
variability in quasars, such as disk-instability (Kawaguchi et al.
1998), gravitational microlensing (Hawkings 1993, 1996), and
starburst (Terlevich et al. 1992). However, all of them are still
far from clear.

Hawkins (1993, 1996) explained the observed AGN variability by
invoking gravitational microlensing. In this model, a quasar's light
is lensed by a large population of compact bodies with
planetary-mass. The microlensing model has two parameters: the
Einstein radius of the lenses and their mean transverse velocity
(Hawkins 2002). However, the two parameters generally can not be
obtained observationally. In addition, microlensing events should be
extremely rare at low redshift (Vanden Berk et al. 2004). This
explanation can be easily excluded for the PG quasars studied in
this paper. We find the quasar's variability is related with the E1
space, which strongly indicates that the variability must be caused
by an intrinsic mechanism rather than an external one.

Someone holds the idea that AGN variability might be caused by a
series of discrete outbursts, such as supernova explosions (Aretxaga
et al. 1997). However, this model can not explain the relationship
between the luminosity and variability amplitude as argued by Pica
\& Smith (1983). Alternatively, Terlevich et al. (1992) explained
the AGN variability as originating from the supernova remnants
(SNRs) occurring in the innermost regions of AGNs. The long-term and
short-term variability observed in AGNs could be explained by the
long-term decay of the SNRs and cooling instability after the onset
of the radiative phase, respectively. The cooling time before the
radiative phase is $\sim$0.6 year, and beyond this, the phase is
reduced to $\sim$6 days.

The disk-instability model is much more popular than the other
models. This model interprets the variability as an occasional flare
event or blob formation caused by the instability in the accretion
disk. Kawaguchi et al. (1998) compared the logarithmic slopes of the
structure function between the disk-instability model and star-burst
model, and their observation of quasar 0957+561 supports the
disk-instability model. Vanden Berk et al. (2004) studied
photometric variability of 25,000 quasars from SDSS, and found that
their results favor the disk-instability model. However, it is still
unclear how changes in the accretion rate or the resulting
luminosity changes would propagate through the accretion disk.
Recently, Li \& Cao (2008) proposed a disk model, and suggested that
the disk temperature change would lead to systematic spectral shape
difference, which could explain the correlation between the
variability and L/$\rm{L_{Edd}}$ or $\rm{M_{BH}}$ discovered by Wold
et al. (2007) and Wilhite et al. (2008).

Although our analysis can not discriminate which model, starburst or
disk-instability, is favored for a quasar's variability, the
extension of the E1 space to the variability implies that either: 1)
the stability of the accretion rate (or gas supply) changes along
the E1 sequence in the disk-instability model, or 2) the intensity
of star formation activity changes along the E1 sequence.

\section {Conclusions}
\label{sect:conclusion}

By studying the variability of the 42 PG quasars monitored by Giveon
et al. (1999), using both direct correlations and PCA analysis, we
find that the E1 space can be extended to include quasar's optical
variability. The link between this variability and Eddington
ratio/black hole mass is discussed, and we propose that the
Eddington ratio may be a key factor in determining the variability
of quasars.

\begin{acknowledgements}

We would thank the anonymous referee for constructive comments that
improved the paper. We also thank Prof. J. M. Wang, Dr. W. H. Bian,
D. W. Xu, J. S. Deng and James Wicker for valuable discussion and
help. This work was supported by the National Science Foundation of
China (grant 10803008 and 10873017) and National Basic Research
Program of China.
\end{acknowledgements}

\clearpage

\begin{figure}
  \centering
  \includegraphics[width=112mm,height=116mm]{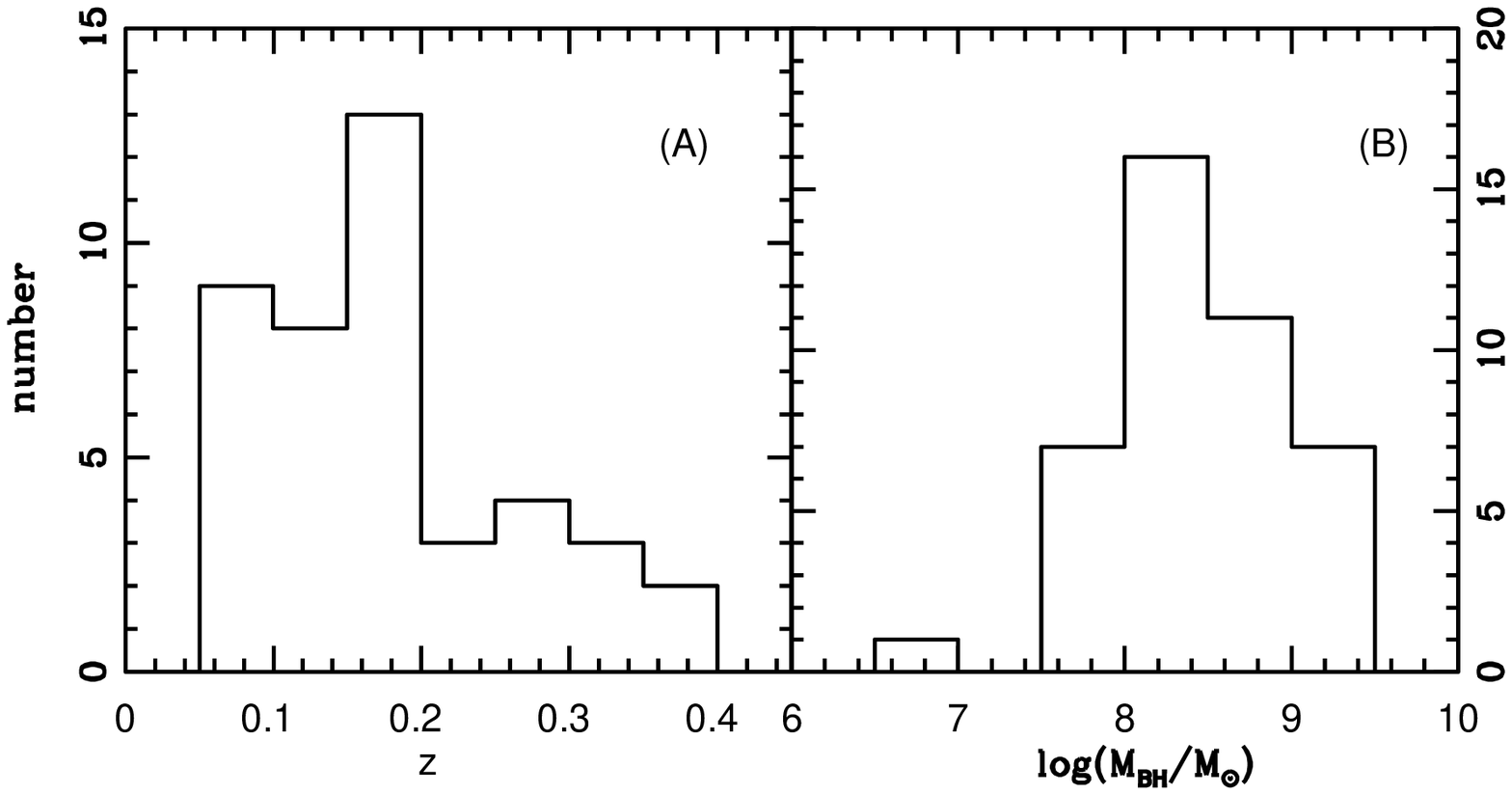}
  \vspace{-5mm}
  \caption{{\small Distribution of redshift (Panel A)
and black hole mass (Panel B) for the 42 PG quasars studied in this
paper. } }
  \label{Fig:fig1}
\end{figure}

\begin{figure}
  \centering
  \includegraphics[width=112mm,height=116mm]{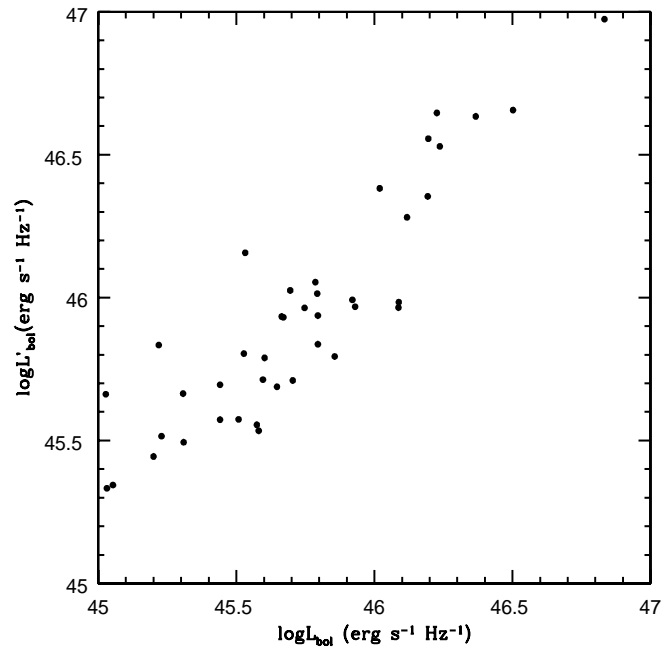}
  \vspace{-5mm}
  \caption{{\small Comparison between the value of $\rm{L_{bol}}$ calculated from
  Equation 2 and 3. The horizon axis represents the $\rm{L_{bol}}$ from Equation 2 (Kaspi et al. 2000),
  and the vertical axis represents the $\rm{L'_{bol}}$ from Equation 3 (Marconi et al. 2004).
  It is clear that the two independent measurements are
  consistent with each other.} }
  \label{Fig:fig2}
\end{figure}

\begin{figure}
  \centering
  \includegraphics[width=112mm,height=116mm]{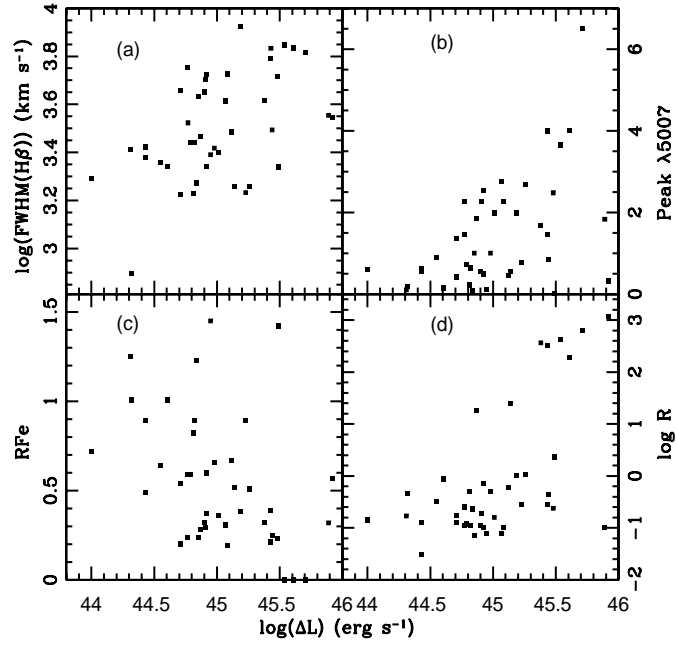}
  \vspace{-5mm}
  \caption{{\small Plots of $\Delta$L vs E1 parameters.
  The horizontal axis represents the $\Delta$L, and the
  vertical axis represents the E1 parameters: (a) log(FWHM(H$\beta$))
   ($\rm{r_s}$=0.450, $\rm{P_s}$=0.004) (b) peak $\lambda$5007
   ($\rm{r_s}$=0.445, $\rm{P_s}$=0.0044). (c) RFeII ($\rm{r_s}$=-0.441,
    $\rm{P_s}$=0.0048). (d) log R($\rm{r_s}$=0.476, $\rm{P_s}$=0.0023).
\label{fig3} } }
  \label{Fig:fig3}
\end{figure}

\begin{figure}
  \centering
  \includegraphics[width=112mm,height=116mm]{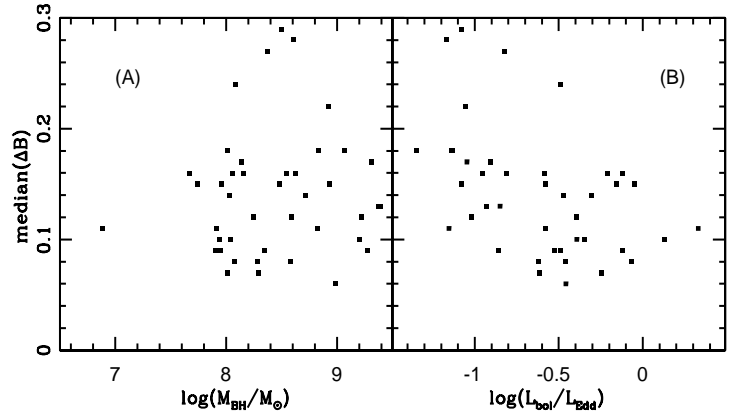}
  \vspace{-5mm}
  \caption{{\small Panel A: median magnitude change vs black hole mass
($r_s = 0.124$, $P_s = 0.4267$). Panel B: median magnitude change vs
Eddington ratio ($r_s = -0.368$, $P_s = 0.0012$). \label{fig4} } }
  \label{Fig:fig4}
\end{figure}

\clearpage

\begin{table}[]
\label{Tab:sample}
\begin{center}

\begin{tabular}{cccccccccc}
\hline\noalign{\smallskip}

Object & z & log($\rm{M_{BH}}$/$\rm{M_{\odot }}$) &
log($\rm{L_{bol}}$/$\rm{L_{Edd}}$) & med($\Delta$B) & log($\Delta$L)
 & FWHM(H$\beta$) & Peak[OIII] & RFe & log R \\
(1) & (2) & (3) & (4) & (5) & (6) & (7) & (8) & (9) & (10)\\
\hline\noalign{\smallskip}
PG 0026+129          &   0.142 & 8.1 & -0.1  & 0.16  &  45.1  &   1860  &   2.68 &    0.51 &   0.03     \\
PG 0052+251          &   0.155 & 8.9 & -1.1  & 0.22  &  45.0  &   5200  &   2.48 &    0.23 &   -0.62    \\
PG 0804+761          &   0.100 & 8.5 & -0.6  & 0.15  &  45.1  &   3070  &   0.46 &    0.67 &   -0.22    \\
PG 0838+770          &   0.131 & 8.2 & -0.6  & 0.16  &  44.7  &   2790  &   0.65 &    0.89 &   -0.96    \\
PG 0844+349          &   0.064 & 7.9 & -0.6  & 0.11  &  44.5  &   2420  &   0.55 &    0.89 &   -1.52    \\
PG 0923+201          &   0.190 & 8.0 & -0.1  & 0.18  &  45.0  &   7610  &   0.60 &    0.72 &   -0.85    \\
PG 0953+414          &   0.239 & 8.7 & -0.5  & 0.14  &  45.4  &   3130  &   0.84 &    0.25 &   -0.36    \\
PG 1001+054          &   0.161 & 7.7 & -0.2  & 0.15  &  44.7  &   1740  &   0.23 &    0.82 &   -0.30    \\
PG 1012+008          &   0.185 & 8.2 & -0.4  & 0.12  &  45.0  &   2640  &   1.00 &    0.66 &   -0.30    \\
PG 1048+342          &   0.167 & 8.4 & -0.8  & 0.27  &  44.7  &   3600  &   1.83 &    0.32 &   -1.00    \\
PG 1100+772          &   0.313 & 9.3 & -0.9  & 0.09  &  45.6  &   6160  &   3.99 &    0.21 &   2.52     \\
PG 1114+445          &   0.144 & 8.6 & -1.0  & 0.12  &  44.7  &   4570  &   1.36 &    0.20 &   -0.89    \\
PG 1115+407          &   0.154 & 7.7 & -0.2  & 0.16  &  44.6  &   1720  &   0.41 &    0.54 &   -0.77    \\
PG 1121+422          &   0.234 & 8.0 & -0.3  & 0.14  &  44.9  &   2220  &   2.55 &    0.37 &   -1.00    \\
PG 1151+117          &   0.176 & 8.5 & -1.0  & 0.16  &  44.8  &   4300  &   1.00 &    0.24 &   -1.15    \\
PG 1202+281          &   0.165 & 8.6 & -1.2  & 0.28  &  44.6  &   5050  &   2.27 &    0.29 &   -0.72    \\
PG 1211+143          &   0.085 & 8.0 & -0.1  & 0.15  &  45.1  &   1860  &   0.55 &    0.52 &   1.39     \\
PG 1226+023          &   0.158 & 9.2 & -0.3  & 0.10  &  46.0  &   3520  &   0.33 &    0.57 &   3.06     \\
PG 1229+204          &   0.064 & 8.1 & -0.9  & 0.17  &  44.4  &   3360  &   1.46 &    0.59 &   -0.96    \\
PG 1307+085          &   0.155 & 8.9 & -1.1  & 0.15  &  45.0  &   2360  &   2.26 &    0.19 &   -1.00    \\
PG 1309+355          &   0.184 & 8.3 & -0.5  & 0.09  &  45.0  &   2940  &   1.86 &    0.28 &   1.26     \\
PG 1322+659          &   0.168 & 8.3 & -0.5  & 0.08  &  45.0  &   2790  &   0.72 &    0.59 &   -0.92    \\
PG 1351+640          &   0.087 & 8.8 & -1.2  & 0.11  &  44.8  &   5660  &   2.27 &    0.24 &   -0.59    \\
PG 1354+213          &   0.300 & 8.6 & -0.8  & 0.16  &  45.0  &   4140  &   2.75 &    0.31 &   -1.10    \\
PG 1402+261          &   0.164 & 7.9 & -0.1  & 0.09  &  45.0  &   1910  &   0.09 &    1.23 &   -0.64    \\
PG 1404+226          &   0.098 & 6.9 &  0.3  & 0.11  &  44.4  &   880   &   0.18 &    1.01 &   -0.33    \\
PG 1411+442          &   0.089 & 8.1 & -0.6  & 0.08  &  44.6  &   2670  &   0.63 &    0.49 &   -0.89    \\
PG 1415+451          &   0.114 & 8.0 & -0.6  & 0.07  &  44.6  &   2620  &   0.10 &    1.25 &   -0.77    \\
PG 1426+015          &   0.086 & 9.1 & -1.3  & 0.18  &  44.9  &   6820  &   1.47 &    0.39 &   -0.55    \\
PG 1427+480          &   0.221 & 8.1 & -0.5  & 0.24  &  44.8  &   2540  &   1.99 &    0.36 &   -0.80    \\
PG 1444+407          &   0.267 & 8.3 & -0.2  & 0.07  &  45.2  &   2480  &   0.12 &    1.45 &   -1.10    \\
PG 1512+370          &   0.371 & 9.4 & -0.9  & 0.13  &  45.6  &   6810  &   4.00 &    0.00 &   2.28     \\
PG 1519+226          &   0.137 & 7.9 & -0.4  & 0.10  &  44.7  &   2220  &   0.16 &    1.01 &   -0.05    \\
PG 1545+210          &   0.266 & 9.3 & -1.0  & 0.17  &  45.4  &   7030  &   3.66 &    0.00 &   2.62     \\
PG 1613+658          &   0.129 & 9.2 & -1.5  & 0.12  &  44.8  &   8450  &   1.99 &    0.38 &   0.00     \\
PG 1617+175          &   0.114 & 8.8 & -1.1  & 0.18  &  44.9  &   5330  &   0.48 &    0.60 &   -0.14    \\
PG 1626+554          &   0.133 & 8.5 & -1.1  & 0.29  &  44.6  &   4490  &   0.56 &    0.32 &   -0.96    \\
PG 1700+518          &   0.292 & 8.6 & -0.1  & 0.08  &  45.7  &   2210  &   0.00 &    1.42 &   0.37     \\
PG 1704+608          &   0.371 & 9.4 & -0.8  & 0.13  &  45.7  &   6560  &   6.50 &    0.00 &   2.81     \\
PG 2130+099          &   0.061 & 7.9 & -0.5  & 0.09  &  44.5  &   2330  &   0.89 &    0.64 &   -0.49    \\
PG 2233+134          &   0.325 & 8.0 &  0.1  & 0.10  &  45.3  &   1740  &   0.77 &    0.89 &   -0.55    \\
PG 2251+113          &   0.323 & 9.0 & -0.5  & 0.06  &  45.7  &   4160  &   1.69 &    0.32 &   2.56     \\

  \noalign{\smallskip}\hline
\end{tabular}
\caption{Column(1): object Name. Column(2): redshift. Column(3): the
logarithm of the black hole mass, as given by Vestergarrd \&
Peterson (2006). Column(4): the logarithm of the Eddington ratio, as
given by Vestergarrd \& Peterson (2006). Column(5):
median($\Delta$B), in units of mag. Column(6): log($\Delta$L),
defined as the Equation 1. Column(7): the FWHM of the broad
component of H$\beta$, in units of km $s^{-1}$. Column(8): the peak
height of [OIII]5007. Column(9): the RFeII. Column(10): the
logarithm of the radio power.}

\end{center}
\end{table}

\clearpage

\begin{table}[]
\caption[]{Spearman Rank-Order Correlation Coefficient of the
correlations shown in Figure 3} \label{Tab:E1 coefficients}
\begin{center}\begin{tabular}{ccccc}
\hline\noalign{\smallskip}

 $\Delta$ L vs & FWHM(H$\beta$) & Peak $\lambda$5007 & RFe & Log R \\
  \hline\noalign{\smallskip}

$r_{s}$ & 0.450 & 0.445 & -0.441 & 0.476\\
$P_{s}$ & 0.004 & 0.004 & 0.005 & 0.002\\

  \noalign{\smallskip}\hline
  \end{tabular}\end{center}
\end{table}

\begin{table}[]
\caption[]{Correlations of Eigenvectors with line and continuum
properties} \label{Tab:E1 coefficients}
\begin{center}\begin{tabular}{ccccc}
\hline\noalign{\smallskip}

 Property  & Eigenvector 1 & Eigenvector 2 & Eigenvector 3 &Eigenvector 4 \\
  \hline\noalign{\smallskip}

Eigenvalue & 3.89 & 1.87 & 1.24 & 1.03\\
Cumulative & 35.3\% & 52.3\% & 63.6\% & 73.0\%\\
\hline\noalign{\smallskip}
log R  & 0.68 & 0.54 & -0.15 & 0.03 \\
$\rm{EW(H\beta_{b})}$  & -0.05 & -0.60 & -0.55 & -0.08\\
R ($\lambda$5007) & 0.82 & 0.04 & 0.43 & -0.13 \\
R ($\lambda$4686) & -0.14 & -0.55 & 0.44 & -0.23 \\
RFe   & -0.79 & 0.45 & -0.03 & -0.04\\
Peak ($\lambda$5007) & 0.92 & -0.07 & 0.24 & -0.12 \\
$\rm{FWHM(H\beta_{b})}$  & 0.68 & -0.21 & -0.32 & 0.34\\
H$\beta$ shift & -0.10 & 0.39 & 0.46 & 0.05\\
H$\beta$ shape & 0.07 & 0.27 & -0.29 & -0.89\\
H$\beta$ asymm & -0.70 & -0.33 & 0.02 & 0.15\\
log ($\Delta$L) & 0.54 & 0.57 & -0.30 & 0.10\\
  \noalign{\smallskip}\hline
  \end{tabular}\end{center}
\end{table}

 \label{lastpage}

\end{document}